\documentclass[aps,prd,twocolumn,superscriptaddress,showpacs]{revtex4}
\usepackage{epsfig,epsf}
\usepackage{amsmath}
\usepackage{amsthm}
\usepackage{amsfonts}
\usepackage{amssymb}
\usepackage{dsfont}
\usepackage{multirow}
\usepackage{appendix}
\usepackage{slashed}
\usepackage[active]{srcltx}
\usepackage{psfrag}

\begin{document}

\title{ Width of the exotic  $X_b(5568)$ state through  its strong decay to $B_s^{0} \pi^{+}$ }
\date{\today}
\author{S.~S.~Agaev}
\affiliation{Department of Physics, Kocaeli University, 41380 Izmit, Turkey}
\affiliation{Institute for Physical Problems, Baku State University, Az--1148 Baku,
Azerbaijan}
\author{K.~Azizi}
\affiliation{Department of Physics, Do\v{g}u\c{s} University, Acibadem-Kadik\"{o}y, 34722
Istanbul, Turkey}
\author{H.~Sundu}
\affiliation{Department of Physics, Kocaeli University, 41380 Izmit, Turkey}

\begin{abstract}
The width of the newly observed exotic state $X_b(5568)$  is calculated via its dominant strong decay to $%
B_s^{0} \pi^{+}$ using the QCD sum rule method on the
light-cone in conjunction with the soft-meson approximation. To this end,
the vertex $X_{b}B_{s}\pi$ is studied and the strong coupling $g_{X_{b}B_{s}\pi}$
is computed employing for $X_b(5568)$ state the interpolating
diquark-antidiquark current of the $[su][\overline{b}\overline{d}]$ type.
The obtained prediction for the decay width of $X_b(5568)$
is confronted and a nice agreement found with the experimental data of the D0 Collaboration.
\end{abstract}

\maketitle

\section{Introduction}
Starting from the discovery of the charmoniumlike resonance $X(3872)$ by
Belle Collaboration \cite{Choi:2003ue},  later confirmed by some other
experiments \cite{Abazov:2004kp,Acosta:2003zx,Aubert:2004ns}, investigation of the exotic
states became one of the interesting directions in  hadron physics. The
exotic states, that is particles inner structure of which cannot be
described within the two- or three quark scheme of standard hadron
spectroscopy, provide rich information to check contemporary theories
claiming to explain variety of new phenomena. They can be produced in
numerous exclusive and inclusive processes icluding the $B$ meson decays
and $\overline{p}p$ collisions. Now, the hadron spectroscopy contains the
numerous family of the exotic states XYZ, discovered and studied during past
years. Relevant experimental investigations are concentrated on measurements of
the masses and decay widths of these states, on exploration of their spins and
parities. At the same time, theoretical studies are aimed to invent new
approaches and methods to calculate parameters of the exotic
particles (see for instance  \cite%
{Swanson:2006st,Klempt:2007cp,Godfrey:2008nc,Voloshin:2007dx,Nielsen:2009uh,
Faccini:2012pj,Esposito:2014rxa,Chen:2016qju} and references therein).

There are various models in the literature suggested to reveal the internal
quark-antiquark structure of the exotic states and explain the wide range of
corresponding experimental data. One of the popular models is the four-quark
or tetraquark picture of the exotic particles. In accordance with this
approach, new charmoniumlike states are composed of two heavy and two light
quarks. These quarks may cluster into the colored diquark and antidiquark,
which are organized in such a way that to reproduce quantum numbers of the
corresponding exotic states. Other possibilities of the organization of the
exotic particles in the context of the tetraquark picture include a
meson-molecule and hadro-quarkonium models. Among alternative views on the
nature of the exotic states one should mention the conventional charmonium  model
and its hybrid extensions. All of these models consider
the exotic states as particles containing a  $c\bar{c}$ component.
In fact, the known exotic states consist of $c\bar{c}$ and two
light quarks. In other words, the number of the quark flavors inside of the
known four-quark exotic particles does not exceed three.

Recently,  the D0 Collaboration in Ref.\ \cite{D0:2016mwd} reported the observation
of a narrow structure $X(5568)$ in the decays $X(5568) \to B_{s}^{0} \pi^{\pm}$, $%
B_{s}^{0} \to J/\psi \phi$, $J/\psi \to \mu^{+} \mu^{-}$, $\phi \to
K^{+}K^{-}$. This result was based on  $p\bar{p}$ data collected at the
Fermilab Tevatron at $\sqrt{s}=1.96\ \mathrm{TeV}$. The mass of the new state
extracted from the experiment equals to $m_{X_{b}}=5567.8 \pm 2.9 \mathrm{%
(stat)}^{+0.9}_{-1.9} \mathrm{(syst)}\, \mathrm{MeV}$, whereas its decay width
was estimated as $\Gamma=21.9 \pm 6.4 \mathrm{(stat)}^{+5.0}_{-2.5} \mathrm{%
(syst)}\, \mathrm{MeV}$. The exotic state  $X(5568)$ is supposedly a scalar particle with
the quantum numbers $J^{PC}=0^{++}$ and built of four distinct quark
flavors. In fact, from the existing decay channel $X(5568) \to B_{s}^{0}
\pi^{\pm}$ one can conclude that the state $X(5568)$ contains the valence $b,\, s, \,
u$ and $d$ quarks. This state can be described as the quark-antiquark
bound state with one of the possible structures $[bu][\bar{d}\bar{s}]$,
$[bd][\bar{s}\bar{u}]$, $[su][\bar{b}\bar{d}]$ and
$[sd][\bar{b}\bar{u}]$, or may be considered as  a  molecule composed of B and $\bar K$
mesons \cite{D0:2016mwd}.

To differ the $X(5568)$ from the conventional members of the X family
of exotic particles, in Ref.\ \cite{Agaev:2016mjb} we introduced the notation
$X_b(5568)$. In the present work we will use this abbreviation, as well.
In Ref.\ \cite{Agaev:2016mjb}, adopting the diquark-antidiquark structure
$X_b=[su][\bar{b}\bar{d}]$ we calculated, for the first time, the mass and
decay constant of the $X_b(5568)$ state. We employed QCD two-point sum rule
method and taken into account the vacuum condensates up to eight dimensions.
Our prediction for the mass of the $X_b(5568)$ state is in a nice agreement
with data of the D0 Collaboration.

The mass and  pole residue of the exotic state $X(5568)$ were also calculated
in Ref.\ \cite{Wang:2016mee}. The mass of the state $X(5568)$ was evaluated in Refs.\  \cite{Wang:2016tsi,Chen:2016mqt,Zanetti:2016wjn}, as well.
To perform relevant  calculations, in these works
different forms of diquark-antidiquark interpolating currents were used.  The obtained results for $m_{X}$
agree with each other, and are consistent with experimental data of the D0 Collaboration.

In the present paper we extend our investigation of the newly observed state
$X_{b}(5568)$ by calculating the width of the dominant decay $X_{b}(5568)\rightarrow
B_{s}^{0}\pi ^{+}$. To this end, we compute the strong coupling $g_{X_{b}B_{s}\pi }$ by applying methods
of QCD light-cone sum rule (LCSR) and soft-meson
approximation \cite{Balitsky:1989ry,Ioffe:1983ju,Belyaev:1994zk}. The soft-meson
approximation is required because the $X_{b}$ state contains the four
valence quarks, and as a result, the light-cone expansion of the correlation
functions reduces to the short-distance expansion in terms of local matrix
elements. This approximation was applied in our previous work Ref.\ \cite{Agaev:2016dev} 
to calculate the decay widths of the $Z_{c}(3900)$ state, where a good agreement with 
the experimental data and available theoretical results was found. 

This paper is structured in the following way. In section \ref{sec:Vertex},
we calculate the strong coupling $g_{X_{b}B_{s}\pi }$ and width of the
decay $X_{b}(5568)\rightarrow B_{s}^{0}\pi ^{\pm }$. Section \ref{sec:Num}
contains our numerical result. Here we compare our prediction for the width
of  the $X_{b}(5568)$ state with the relevant experimental data of the D0
Collaboration. This section contains also our concluding remarks.

\section{THE vertex $X_bB_s\protect\pi $ and strong decay $X_b \to
B_s^{0} \protect\pi^{+} $}

\label{sec:Vertex}

In this section we calculate the width of the $X_b \to B_s^{0}
\pi $ decay. To this purpose, as the first step we calculate the strong coupling
$g_{X_bB_s \pi }$ by means of the QCD light-cone sum rules method applying 
the soft-meson approximation. In order to get the sum rule
expression for the coupling $g_{X_bB_s \pi }$ we consider the
correlation function
\begin{equation}
\Pi(p,q)=i\int d^{4}xe^{ipx}\langle \pi (q)|\mathcal{T}%
\{J^{B_{s}}(x)J^{X_{b}\dag }(0)\}|0\rangle ,  \label{eq:CorrF3}
\end{equation}
where the interpolating currents are given as
\begin{equation}
J^{X_{b}}(x)=\varepsilon ^{ijk}\varepsilon ^{imn}\left[ s^{j}(x)C\gamma
_{\mu }u^{k}(x)\right] \left[ \overline{b}^{m}(x)\gamma ^{\mu }C\overline{d}%
^{n}(x)\right],  \label{eq:Diq}
\end{equation}
and
\begin{equation}
J^{B_s}(x)=\overline{b}_{l}(x)i\gamma _{5 }s_{l}(x).  \label{eq:Bcur}
\end{equation}%
In Eqs.\ (\ref{eq:Diq}) and (\ref{eq:Bcur}) $i, j, k, m, n$ and $l$ are the color
indices and $C$ is the charge conjugation matrix.

In terms of the physical degrees of freedom the correlation function $\Pi (p,q)$ is determined by the expression 
\begin{eqnarray}
\Pi ^{\mathrm{Phys}}(p,q) &=&\frac{\langle 0|J^{B_{s}}|B_{s}\left(
p\right) \rangle }{p^{2}-m_{B_{s}}^{2}}\langle B_{s}^{0}\left( p\right) \pi
(q)|X_{b}(p^{\prime })\rangle  \notag \\
&&\times \frac{\langle X_{b}(p^{\prime })|J^{X_{b}\dagger }|0\rangle }{%
p^{\prime 2}-m_{X_{b}}^{2}}+\ldots ,  \label{eq:CorrF4}
\end{eqnarray}%
where by dots we denote contributions of the higher resonances and
continuum states. Here $p$, $q$ and $p^{\prime }=p+q$, are the momenta of $%
B_{s}^{0}$, $\pi $, and the $X_{b}$ states, respectively. To compute the correlation
function we introduce also the matrix elements
\begin{eqnarray}
&&\langle 0|J^{B_{s}}|B_{s}\left( p\right) \rangle =\frac{%
f_{B_{s}}m_{B_{s}}^{2}}{m_{b}+m_{s}},  \notag \\
&&\langle X_{b}(p^{\prime })|J^{X_{b}\dagger }|0\rangle =f_{X_{b}}m_{X_{b}},
\notag \\
&&\langle B_{s}^{0}\left( p\right) \pi (q)|X_{b}(p^{\prime })\rangle
=g_{X_{b}B_{s}\pi }p\cdot p^{\prime }.  \label{eq:Mel}
\end{eqnarray}
In Eq.\ (\ref{eq:Mel}) by $m_{X_{b}}$ and $f_{X_{b}}$ we denote the mass and decay
constant of the $X_{b}$ state, whereas $m_{B_{s}}$ and $f_{B_{s}}$ are
the same parameters of the $B_{s}^{0}$ meson.

Using these matrix elements for the correlation function we obtain
\begin{equation}
\Pi ^{\mathrm{Phys}}(p,q)=\frac{%
f_{B_{s}}f_{X_{b}}m_{X_{b}}m_{B_{s}}^{2}g_{X_{b}B_{s}\pi }}{\left( p^{\prime
2}-m_{X_{b}}^{2}\right) \left( p^{2}-m_{B_{s}}^{2}\right) (m_{b}+m_{s})}%
p\cdot p^{\prime}.  \label{eq:CorrF5}
\end{equation}%
In the soft-meson limit accepted in the present work $q=0$, and as a result, $%
p=p^{\prime }$. The reason why we apply the soft-meson limit was explained
in rather detailed form in our previous article \cite{Agaev:2016dev}.
Nevertheless, for completeness we provide briefly corresponding arguments.
In fact, the $X_{b} $ state and interpolating current Eq.\ (\ref{eq:Diq})
contain four quark fields at the same space-time location. Substitution of
this current into the correlation function and subsequent contraction of the
$b$ and $s$ quark fields yield expressions, where the remaining light quarks
are placed between the $\pi $ meson and vacuum states forming local
matrix elements. Stated differently, we appear in the situation when
dependence of the correlation function on the meson distribution amplitudes
disappears and integrals over the meson DAs reduce to overall normalization
factors. In the context of the QCD LCSR method such situation is possible in
the kinematical limit $q\rightarrow 0$, when the light-cone expansion is
replaced by the short-distant one. As a result, instead of the expansion in
terms of DAs, one gets expansion over the local matrix elements \cite%
{Belyaev:1994zk}. In this limit the relevant invariant
amplitudes in the correlation function depend only on the variable $p^{2}$.

In the case under consideration the corresponding invariant amplitude reads
\begin{eqnarray}
&&\Pi ^{\mathrm{Phys}}(p^{2})=\frac{%
f_{B_{s}}f_{X_{b}}m_{X_{b}}m_{B_{s}}^{2}g_{X_{b}B_{s}\pi }}{\left(
p^{2}-m_{X_{b}}^{2}\right) \left( p^{2}-m_{B_{s}}^{2}\right) (m_{b}+m_{s})}%
m^{2}  \notag \\
&&+\Pi ^{\mathrm{(RS:C)}}(p^{2}),  \label{eq:CorrF5A}
\end{eqnarray}%
where $m^{2}=\left( m_{X_{b}}^{2}+m_{B_{s}}^{2}\right) /2.$ In Eq.\ (\ref{eq:CorrF5A}),  $\Pi ^{%
\mathrm{(RS:C)}}(p^{2})$ is the contribution arising from the higher
resonances and continuum states.

What is also important, instead of the two-variable Borel transformation on $%
p^{2}$ and $p^{\prime 2}$ ,\ now we have to use the one-variable Borel
transformation on $p^{2}$: this fact plays a crucial role in deriving sum
rules for the strong couplings. Indeed, the soft-meson approximation
considerably simplifies the QCD side of the sum rules, but leads to more
complicated expression for its hadronic representation. In the soft  limit,
the ground state contribution can be written in the form
\begin{equation}
\Pi ^{\mathrm{Phys}}(p^{2})\cong \frac{%
f_{B_{s}}f_{X_{b}}m_{X_{b}}m_{B_{s}}^{2}g_{X_{b}B_{s}\pi }}{\left(
p^{2}-m^{2}\right) ^{2}(m_{b}+m_{s})}m^{2}.
\end{equation}%
The Borel transformation on the variable $p^{2}$ applied to this
correlation function yields
\begin{eqnarray}
&&\Pi ^{\mathrm{Phys}}(M^{2})=\frac{%
f_{B_{s}}f_{X_{b}}m_{X_{b}}m_{B_{s}}^{2}g_{X_{b}B_{s}\pi }}{(m_{b}+m_{s})}%
m^{2}  \notag \\
&&\times \frac{1}{M^{2}}e^{-m^{2}/M^{2}}.
\end{eqnarray}%
In the soft-meson limit we have to use the one-variable Borel transformation, therefore
transitions from the exited states $m^{\ast }>m_{B_{s}}$ to the ground state in the
$B_{s}^{0}$ channel (similar arguments are valid for the $X_{b}$ channel, as
well) contribute to the hadronic part of the sum rules. The relevant contributions 
even after the Borel transformation are not suppressed relative to the ground state one  \cite{Belyaev:1994zk,Ioffe:1983ju}. To remove from the sum rules
unsuppressed contributions we employ  a
prescription elaborated in Ref.\ \cite{Ioffe:1983ju} and apply  the operator
\begin{equation}
\left( 1-M^{2}\frac{d}{dM^{2}}\right) M^{2}e^{m^{2}/M^{2}}  \label{eq:softop}
\end{equation}%
to both sides of the sum rule expression.

To find the QCD side of the sum rules one should calculate the correlation function 
in the quark-gluon degrees of freedom. Contraction of the $s$ and $b$-quark fields results in 
the expression
\begin{eqnarray}
&&\Pi ^{\mathrm{QCD}}(p,q)=-\int d^{4}xe^{ipx}\varepsilon ^{ijk}\varepsilon
^{imn}\left[ \gamma ^{\mu }\widetilde{S}_{s}^{lj}(x){}\gamma _{5}\right.
\notag \\
&&\left. \times \widetilde{S}_{b}^{ml}(-x){}\gamma _{\mu }\right] _{\alpha
\beta }\langle \pi (q)|\overline{u}_{\alpha }^{k}(0)d_{\beta
}^{n}(0)|0\rangle ,  \label{eq:CorrF6}
\end{eqnarray}%
where $\alpha $ and $\beta $ are the spinor indices. In Eq.\ (\ref{eq:CorrF6}%
) we introduce the notation
\begin{equation*}
\widetilde{S}_{b(s)}^{ij}(x)=CS_{b(s)}^{ijT}(x)C,
\end{equation*}%
with $S_{s}^{ij}(x)$ and $S_{b}^{ij}(x)$ being the $s$- and $b$-quark
propagators, respectively.

In general, for calculation of the $\Pi ^{\mathrm{QCD}}(p,q)$ we have to use
the light-cone expansion for the $s$- and $b$-quark propagators. But because
in the matrix elements the light quark fields are already fixed at the point
$x=0$, it is enough in calculations to utilize the local propagators. We
choose the $s$-quark propagator $S_{q}^{ij}(x)$ in the $x$-space in the form%
\begin{eqnarray}
&&S_{q}^{ij}(x)=i\delta _{ij}\frac{\slashed x}{2\pi ^{2}x^{4}}-\delta _{ij}%
\frac{m_{s}}{4\pi ^{2}x^{2}}-\delta _{ij}\frac{\langle \overline{s}s\rangle
}{12}  \notag \\
&&+i\delta _{ij}\frac{\slashed xm_{s}\langle \overline{s}s\rangle }{48}%
-\delta _{ij}\frac{x^{2}}{192}\langle \overline{s}g\sigma Gs\rangle +i\delta
_{ij}\frac{x^{2}\slashed xm_{s}}{1152}\langle \overline{s}g\sigma Gs\rangle
\notag \\
&&-i\frac{gG_{ij}^{\alpha \beta }}{32\pi ^{2}x^{2}}\left[ \slashed x{\sigma
_{\alpha \beta }+\sigma _{\alpha \beta }}\slashed x\right] +\ldots
\label{eq:qprop}
\end{eqnarray}
For the $b$-quark propagator $S_{b}^{ij}(x)$ we employ the expression \cite%
{Reinders:1984sr}
\begin{eqnarray}
&&S_{b}^{ij}(x)=i\int \frac{d^{4}k}{(2\pi )^{4}}e^{-ikx}\left[ \frac{\delta
_{ij}\left( {\slashed k}+m_{b}\right) }{k^{2}-m_{b}^{2}}\right.  \notag \\
&&\left. -\frac{gG_{ij}^{\alpha \beta }}{4}\frac{\sigma _{\alpha \beta
}\left( {\slashed k}+m_{b}\right) +\left( {\slashed k}+m_{b}\right) \sigma
_{\alpha \beta }}{(k^{2}-m_{b}^{2})^{2}}\right] +\ldots  \label{eq:Qprop}
\end{eqnarray}%
In Eqs.\ (\ref{eq:qprop}) and (\ref{eq:Qprop}) the short-hand notation
\begin{equation*}
G_{ij}^{\alpha \beta }\equiv G_{A}^{\alpha \beta
}t_{ij}^{A},\,\,\,\,A=1,\,2\,\ldots 8,
\end{equation*}%
is used, where $i,\,j$ are color indices, and $t^{A}=\lambda ^{A}/2$ with $%
\lambda ^{A}$ being the standard Gell-Mann matrices. In the nonperturbative
terms the gluon field strength tensor $G_{\alpha \beta }^{A}\equiv G_{\alpha
\beta }^{A}(0)$ is fixed at $x=0.$

To proceed we use the expansion
\begin{equation}
\overline{u}_{\alpha }^{k}d_{\beta }^{m}\rightarrow \frac{1}{4}\Gamma
_{\beta \alpha }^{j}\left( \overline{u}^{k}\Gamma ^{j}d^{m}\right) ,
\label{eq:MatEx}
\end{equation}%
where $\Gamma ^{j}$ is the full set of Dirac matrixes
\begin{equation*}
\Gamma ^{j}=\mathbf{1,\ }\gamma _{5},\ \gamma _{\lambda },\ i\gamma
_{5}\gamma _{\lambda },\ \sigma _{\lambda \rho }/\sqrt{2}.
\end{equation*}%
In order to fix the local matrix elements necessary in our calculations, 
first we consider the perturbative component of the $b$-quark ($\sim \delta _{ml}$) and terms $%
\sim \delta _{lj}$ from the $s$-quark propagators and perform the summation
over the color indices. To this end, we use the overall color factor $\varepsilon
^{ijk}\varepsilon ^{imn}$, color factors of the propagators, and the
projector onto a color-singlet state $\delta ^{km}/3$. As a result, we find that for such
terms the replacement
\begin{equation}
\frac{1}{4}\Gamma _{\beta \alpha }^{j}\left( \overline{u}^{k}\Gamma
^{j}d^{m}\right) \rightarrow \frac{1}{2}\Gamma _{\beta \alpha }^{j}\left(
\overline{u}\Gamma ^{j}d\right)
\end{equation}%
has to be implemented. In the case of the nonperturbative contributions, which appear  as a product of
the perturbative part of one propagator with the $\sim G$ component of the another propagator, 
for example, we obtain
\begin{equation*}
\varepsilon ^{ijk}\varepsilon ^{imn}\delta _{lj}G_{ml}^{\rho \delta }\frac{1%
}{4}\Gamma _{\beta \alpha }^{r}\left( \overline{u}^{k}\Gamma
^{r}d^{m}\right) \rightarrow -\frac{1}{4}\Gamma _{\beta \alpha }^{r}\left(
\overline{u}\Gamma ^{r}G^{\rho \delta }d\right) .
\end{equation*}%
This recipe enables us to insert  the gluon field
strength tensor $G$ into quark matrix elements, and leads to generation of three-particle local
matrix elements of the pion. We neglect the terms $\sim G^{2}$
in our computations.

Having finished a color summation one can calculate the traces over spinor
indices and perform integrations to extract the imaginary part of the
correlation function $\Pi ^{\mathrm{QCD}}(p,q)$ in accordance with
procedures described in Ref.\ \cite{Agaev:2016dev}. Omitting the technical
details we provide the final expression for the spectral density, which consists
of the perturbative and nonperturbative components
\begin{equation}
\rho ^{\mathrm{QCD}}(s)=\rho ^{\mathrm{pert.}}(s)+\rho ^{\mathrm{n.-pert.}%
}(s).  \label{eq:SD}
\end{equation}%
Calculations show that,  the local matrix element of
the pion, which contributes in the soft-meson limit to the $\mathrm{Im}\Pi ^{\mathrm{QCD}}(p)$ is
\begin{equation}
\langle 0|\overline{d}(0)i\gamma _{5}u(0)|\pi (q)\rangle =f_{\pi }\mu _{\pi
}.  \label{eq:MatE2}
\end{equation}%
Here
\begin{equation}
\mu _{\pi }=\frac{m_{\pi }^{2}}{m_{u}+m_{d}}=-\frac{2\langle \overline{q}%
q\rangle }{f_{\pi }^{2}},  \label{eq:PionEl}
\end{equation}%
where the second equality is the relation between $%
m_{\pi }$, $f_{\pi }$, the quark masses and the quark condensate $\langle
\overline{q}q\rangle $, which follows from the partial conservation of axial
vector current (PCAC).

The components of the spectral density are given by the formulas:
\begin{equation}
\rho ^{\mathrm{pert.}}(s)=\frac{f_{\pi }\mu _{\pi }}{4\pi ^{2}s}\sqrt{%
s(s-4m_{b}^{2})}\left( s+2m_{b}m_{s}-2m_{b}^{2}\right) ,  \label{eq:SD1}
\end{equation}%
and
\begin{eqnarray}
&&\rho ^{\mathrm{n.-pert.}}(s)=\frac{f_{\pi }\mu _{\pi }}{18}\left\{
6\langle \overline{s}s\rangle \left[ -2m_{b}\delta (s-m_{b}^{2})\right.
\right.   \notag \\
&&\left. +sm_{s}\delta ^{^{(1)}}(s-m_{b}^{2})\right] +\langle \overline{s}%
g\sigma Gs\rangle \left[ 6(m_{b}-m_{s})\delta ^{^{(1)}}(s-m_{b}^{2})\right.
\notag \\
&&\left. \left. -3s(m_{b}-2m_{s})\delta ^{(2)}(s-m_{b}^{2})-s^{2}m_{s}\delta
^{(3)}(s-m_{b}^{2})\right] \right\} .  \label{eq:SD2}
\end{eqnarray}%
In Eq.\ (\ref{eq:SD2}) $\delta ^{(n)}(s-m_{b}^{2})=(d/ds)^{n}\delta
(s-m_{b}^{2})$ that appear in extracting the imaginary part of the pole
terms and stem from the well known formula
\begin{equation*}
\frac{1}{s-m_{b}^{2}}=\mathrm{P.V.}\frac{1}{s-m_{b}^{2}}+i\pi \delta
(s-m_{b}^{2}).
\end{equation*}

The continuum subtraction is performed in a standard manner after $\rho
^{h}(s)\rightarrow \rho ^{\mathrm{QCD}}(s)$ replacement. Then, the final sum
rule to evaluate the strong coupling reads
\begin{eqnarray}
&&g_{X_{b}B_{s}\pi }=\frac{(m_{b}+m_{s})}{%
f_{B_{s}}f_{X_{b}}m_{X_{b}}m_{B_{s}}^{2}m^{2}}\left( 1-M^{2}\frac{d}{dM^{2}}%
\right) M^{2}  \notag \\
&&\times \int_{(m_{b}+m_{s})^{2}}^{s_{0}}dse^{(m^{2}-s)/M^{2}}\rho ^{\mathrm{%
QCD}}(s).  \label{eq:SRules}
\end{eqnarray}

The width of the decay $X_{b}\rightarrow B_{s}^{0}\pi ^{+}$ can be found
applying the usual prescriptions and definitions for the strong coupling
together with other matrix elements from Eq.\ (\ref{eq:Mel}) as well as 
the parameters of the $X_{b}$ state. The calculations give%
\begin{eqnarray}
&&\Gamma \left( X_{b}\rightarrow B_{s}^{0}\pi ^{+}\right) =\frac{%
g_{X_{b}B_{s}\pi }^{2}m_{B_{s}}^{2}}{24\pi }\lambda \left( m_{X_{b}},\
m_{B_{s}},m_{\pi }\right)   \notag \\
&&\times \left[ 1+\frac{\lambda ^{2}\left( m_{X_{b}},\ m_{B_{s}},m_{\pi
}\right) }{m_{B_{s}}^{2}}\right] ,  \label{eq:DW}
\end{eqnarray}%
where
\begin{equation*}
\lambda (a,\ b,\ c)=\frac{\sqrt{a^{4}+b^{4}+c^{4}-2\left(
a^{2}b^{2}+a^{2}c^{2}+b^{2}c^{2}\right) }}{2a}.
\end{equation*}%
Equations\ (\ref{eq:SRules}) and (\ref{eq:DW}) are final expressions that
will be used for numerical analysis of the decay channel $X_{b}\rightarrow
B_{s}^{0}\pi ^{+}$.


\section{Numerical results and conclusions}

\label{sec:Num}
\begin{table}[tbp]
\begin{tabular}{|c|c|}
\hline\hline
Parameters & Values \\ \hline\hline
$m_{B_s}$ & $(5366.77\pm0.24)~\mathrm{MeV}$ \\
$f_{B_s}$ & $(222\pm 12)~\mathrm{MeV}$ \\
$m_{b}$ & $(4.18\pm0.03)~\mathrm{GeV}$ \\
$m_{s} $ & $(95 \pm 5)~\mathrm{MeV} $ \\
$\langle \bar{q}q \rangle $ & $(-0.24\pm 0.01)^3$ $\mathrm{GeV}^3$ \\
$\langle \bar{s}s \rangle $ & $0.8\ \langle \bar{q}q \rangle$ \\
$m_{0}^2 $ & $(0.8\pm0.1)$ $\mathrm{GeV}^2$ \\
$\langle \overline{q}g\sigma Gq\rangle $ & $m_{0}^2\langle \bar{q}q \rangle$
\\ \hline\hline
\end{tabular}%
\caption{Input parameters used in calculations.}
\label{tab:Param}
\end{table}

The QCD sum rule for the strong coupling $g_{X_{b}B_{s}\pi}$ and decay width
$\Gamma \left( X_{b}\rightarrow B_{s}^{0}\pi ^{+}\right)$ contain various
parameters that should be fixed in accordance with the standard procedures:
for numerical computations we need the masses and decay constants of the $%
X_b $ and $B_{s}^{0}$ mesons as well as values of the quark and mixed
condensates. In addition to that, QCD sum rules depend on the $b$ and $s$
quark masses. The values of some used parameters are moved to Table \ref%
{tab:Param}. In the calculations we also employ the QCD sum rule predictions for the mass and decay constant of $X_b $
state obtained in our work Ref.\ \cite{Agaev:2016mjb}. The value of the decay constant $f_{B_s}$ is borrowed
from Ref.\ \cite{Baker:2013mwa}.

\begin{figure}[tbp]
\centerline{
\begin{picture}(210,170)(0,0)
\put(-10,0){\epsfxsize8.2cm\epsffile{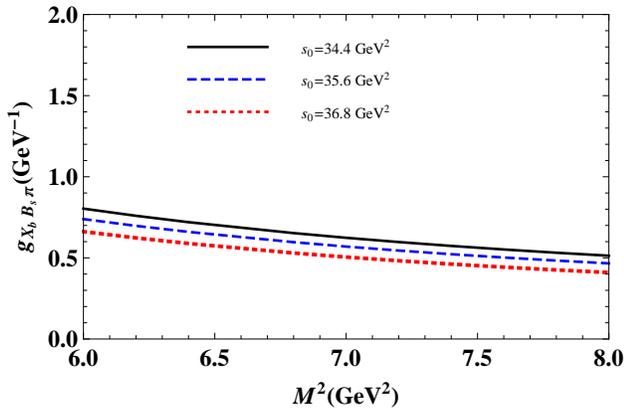}}
\end{picture}
}
\caption{The strong coupling constant $g_{X_{b}B_{s}\pi}$ vs Borel parameter $M^2$ at different fixed values of $s_0$.}
\label{fig:1}
\end{figure}

\begin{figure}
\centerline{
\begin{picture}(200,170)(0,0)
\put(-10,20){\epsfxsize8.2cm\epsffile{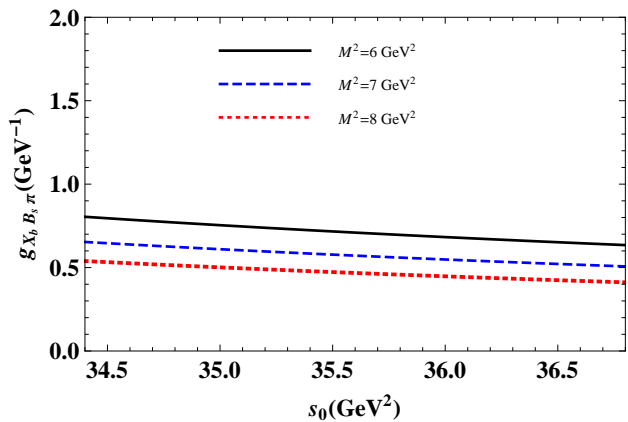}}
\end{picture}}
\caption{The strong coupling constant $g_{X_{b}B_{s}\pi}$ as a function of the  threshold parameter $s_0$ at different fixed values of $M^2$.}
\label{fig:2}
\end{figure}

Calculations also require fixing of the auxiliary parameters, namely the continuum threshold  $s_{0}$ and Borel parameter $M^2$.
 The standard criteria accepted in the sum rule calculations require the practical independence of the physical quantities on these auxiliary parameters.
 The continuum threshold is not totally arbitrary but, in principle, depends on the energy of the first excited state with the same quantum numbers and structure as the particle under
consideration. In the lack of information on the mass of the first excited state in this channel, we follow the 
traditional prescriptions and choose $s_0$ in the interval $(m_{X_{b}}+0.3)^2\ \mathrm{GeV}^2\leq s_0\leq (m_{X_{b}}+0.5)^2\ \mathrm{GeV}^2$, i.e.
\begin{equation}
34.4\,\,\mathrm{GeV}^2 \leq s_{0}\leq 36.8 \,\,\mathrm{GeV}^2.
\end{equation}
 To determine the working window for the Borel parameter,  we demand  not only sufficient suppression of the contributions due to the higher states and continuum,
 but also exceeding of the perturbative contributions over the non-perturbative ones as well as convergence of the OPE series. Technically, the upper limit on $M^2$ is obtained by the requirement
\begin{eqnarray}
\label{nolabel}
\frac{ \int_{0}^{s_0}ds  \rho^{QCD}(s) e^{-s/M^2} }{
 \int_{0}^\infty ds \rho^{QCD}(s) e^{-s/M^2}} ~~>~~ 1/2.
\end{eqnarray}
The lower limit on $M^{2}$ is found by requiring that the perturbative contribution
exceeds the non-perturbative one, and that the higher dimensional terms constitute less than 10\% of the total contribution. 
These requirements lead to the working interval  $6\ \mathrm{GeV}^2\leq M^2\leq 8\ \mathrm{GeV}^2$  for the Borel parameter. Considering these  regions, we
plot the strong coupling constant $g_{X_{b}B_{s}\pi}$ as functions of $M^2$ and $s_0$ in Figs. \ref{fig:1} and \ref{fig:2}. From these figures we see that the coupling $g_{X_{b}B_{s}\pi}$ demonstrates weak
dependence on the Borel and threshold parameters in the selected working regions.

Extracted from the numerical calculations,   value of the strong coupling $g_{X_{b}B_{s}\pi}$ is obtained as
\begin{equation}
g_{X_{b}B_{s}\pi}=(0.60 \pm 0.23)\ \mathrm{GeV}^{-1}.
\end{equation}
For the width of the decay $X_b(5568) \to B_{s}^{0} \pi^{+}$ we get
\begin{equation}
\Gamma(X_{b}\to B_{s}^{0}\pi^{+})=(22.4 \pm 9.2)\ \mathrm{MeV},
\end{equation}
which is in a good consistency with the experimental data of the D0 Collaboration.

In this work we have continued our studies of the newly discovered exotic
state $X_b(5568)$ and computed the width of the strong decay $X_b \to B_{s}^{0}\pi^{+}$
using methods of QCD sum rules on the light-cone and soft-meson approximation.
To this end, first we  found the strong coupling $g_{X_{b}B_{s}\pi}$ that allowed us to
calculate $\Gamma(X_{b}\to B_{s}^{0}\pi^{+})$. Our finding is consistent with the experimental
data of the D0 Collaboration.

\section*{ACKNOWLEDGEMENTS}

The work of S.~S.~A. was supported by the TUBITAK grant 2221-"Fellowship
Program For Visiting Scientists and Scientists on Sabbatical Leave". This
work was also supported in part by TUBITAK under the grant no: 115F183.

\end{document}